\documentclass[aps,prl,twocolumn,reprint]{revtex4-1}	
\usepackage{graphicx,color}

\renewcommand{\vec}[1]{\mathbf{#1}}

\usepackage{dcolumn}
\usepackage{bm}
\usepackage{amsmath}
\usepackage{ulem}

\begin{document}

\title{Non-magnetic ground state of Ni adatoms on Te-terminated bismuth chalcogenide topological insulators}

\author{M.~\surname{Vondr\'a\v{c}ek}} 
\email[E-mail: ]{vondrac@fzu.cz}
\affiliation{Institute of Physics of the Czech Academy of Sciences, Na Slovance~2, CZ-182 21 Prague, Czech Republic}

\author{J.~Honolka}
\affiliation{Institute of Physics of the Czech Academy of Sciences, Na Slovance~2, CZ-182 21 Prague, Czech Republic}

\author{L.~Cornils}
\affiliation{Department of Physics, University of Hamburg, Jungiusstr.~11, DE-20355 Hamburg, Germany}

\author{J.~Warmuth}
\affiliation{Department of Physics, University of Hamburg, Jungiusstr.~11, DE-20355 Hamburg, Germany}

\author{L.~Zhou}
\altaffiliation{Max Planck Institute for Solid State Research, D-70569 Stuttgart, Germany}
\affiliation{Department of Physics, University of Hamburg, Jungiusstr.~11, DE-20355 Hamburg, Germany}

\author{A.~Kamlapure}
\affiliation{Department of Physics, University of Hamburg, Jungiusstr.~11, DE-20355 Hamburg, Germany}

\author{A.~A.~Khajetoorians}
\altaffiliation{Institute for Molecules and Materials (IMM), Radboud University, 6525 AJ Nijmegen, The Netherlands}
\affiliation{Department of Physics, University of Hamburg, Jungiusstr.~11, DE-20355 Hamburg, Germany}

\author{R.~Wiesendanger}
\affiliation{Department of Physics, University of Hamburg, Jungiusstr.~11, DE-20355 Hamburg, Germany}

\author{J.~Wiebe}
\affiliation{Department of Physics, University of Hamburg, Jungiusstr.~11, DE-20355 Hamburg, Germany}

\author{M.~Michiardi}
\affiliation{Department of Physics and Astronomy, Interdisciplinary Nanoscience Center, Aarhus University, DK-8000 Aarhus C, Denmark}

\author{M.~Bianchi}
\affiliation{Department of Physics and Astronomy, Interdisciplinary Nanoscience Center, Aarhus University, DK-8000 Aarhus C, Denmark}

\author{J.~Miwa}
\affiliation{Department of Physics and Astronomy, Interdisciplinary Nanoscience Center, Aarhus University, DK-8000 Aarhus C, Denmark}

\author{L.~Barreto}
\altaffiliation{Centro de Ci\^{e}ncias Naturais e Humanas, Universidade Federal do ABC, Santo Andr\'{e} 09210-580, SP, Brazil}
\affiliation{Department of Physics and Astronomy, Interdisciplinary Nanoscience Center, Aarhus University, DK-8000 Aarhus C, Denmark}

\author{P.~Hofmann}
\affiliation{Department of Physics and Astronomy, Interdisciplinary Nanoscience Center, Aarhus University, DK-8000 Aarhus C, Denmark}

\author{C.~Piamonteze}
\affiliation{Swiss Light Source, Paul Scherrer Institut, CH-5232 Villigen PSI, Switzerland}

\author{J.~Min\'ar}
\affiliation{Department of Chemistry, University of Munich, Butenandtstrasse 5-13, D-81377 M\"unchen, Germany},
\affiliation{New Technologies Research Center, University of West Bohemia, Univerzitn\'i 8, CZ-306 14 Plze\v{n}, Czech Republic}

\author{S.~Mankovsky}
\affiliation{Department of Chemistry, University of Munich, Butenandtstrasse 5-13, D-81377 M\"unchen, Germany}

\author{St.~Borek}
\affiliation{Department of Chemistry, University of Munich, Butenandtstrasse 5-13, D-81377 M\"unchen, Germany}

\author{H.~Ebert}
\affiliation{Department of Chemistry, University of Munich, Butenandtstrasse 5-13, D-81377 M\"unchen, Germany}

\author{M.~Sch\"uler}
\affiliation{Institute for Theoretical Physics, Bremen Center for Computational Material Science, University of Bremen, D-28359 Bremen, Germany}

\author{T.~Wehling}
\affiliation{Institute for Theoretical Physics, Bremen Center for Computational Material Science, University of Bremen, D-28359 Bremen, Germany}

\author{J.-L.~Mi}
\altaffiliation{Institute for Advanced Materials, School of Materials Science and Engineering, Jiangsu University, Zhenjiang 212013, China}
\affiliation{Center for Materials Crystallography, Department of Chemistry and iNANO, Aarhus University, Langelandsgade 140, DK-8000 Aarhus C, Denmark}

\author{B.-B.~Iversen}
\affiliation{Center for Materials Crystallography, Department of Chemistry and iNANO, Aarhus University, Langelandsgade 140, DK-8000 Aarhus C, Denmark}

\date{\today}

\begin{abstract}
We report on the quenching of single Ni adatom moments on Te-terminated Bi$_2$Te$_2$Se and Bi$_2$Te$_3$ topological insulator surfaces. The effect becomes manifested as a missing X-ray magnetic circular dichroism for resonant L$_{3,2}$ transitions into partially filled Ni 3$d$ states of occupancy $n_d = 9.2$. On the basis of a comparative study of Ni and Fe using scanning tunneling microscopy and {\it ab initio} calculations we are able to relate the element specific moment formation to a local Stoner criterion. While Fe adatoms form large spin moments of $m_s=2.54\,\mu_B$ with out-of-plane anisotropy due to a sufficiently large density of states at the Fermi energy, Ni remains well below an effective Stoner threshold for local moment formation. With the Fermi level remaining in the bulk band gap after adatom deposition, non-magnetic Ni and preferentially out-of-plane oriented magnetic Fe with similar structural properties on  Bi$_2$Te$_2$Se surfaces constitute a perfect platform to study off-on effects of time-reversal symmetry breaking on topological surface states.
\end{abstract}


\maketitle

\begin{figure}
\center 
\includegraphics[width=75mm]{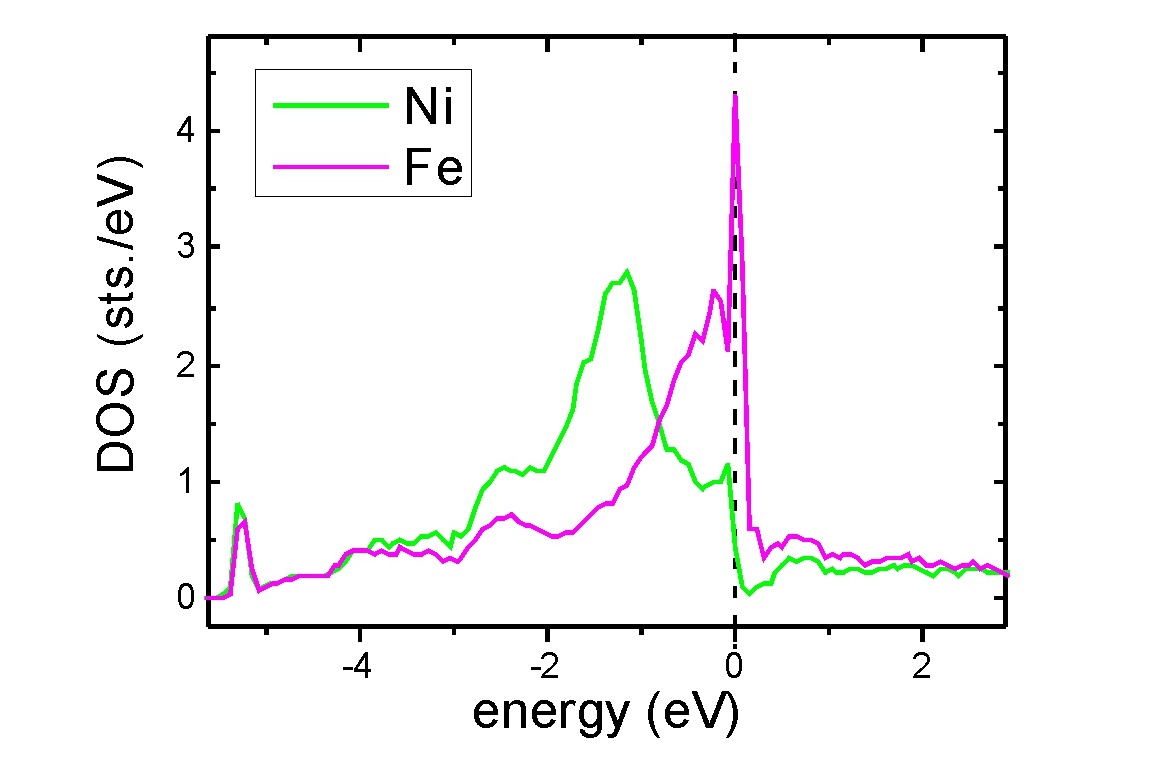}
\caption{Partial densities of states of Ni and Fe adatoms on $\mathrm{Bi_2Te_3}$ calculated for the non-magnetic state.
}
\label{DOS}
\end{figure}

Three-dimensional topological insulators (TI) are insulating in the bulk but host conductive topological surface states (TSSs), which follow a Dirac-like dispersion relation with spin and orbital momentum degrees of freedom chirally locked at a right-angle~\cite{HasanOverview, Qi2011}. As a consequence, currents for momenta $+\vec{k}$ and $-\vec{k}$ carry opposite spin, and 180$^{\circ}$ backscattering is suppressed. While this particular topological spin texture is protected by time-reversal symmetry immanent in TIs, it is vulnerable to perturbations which are odd under time reversal operation.
The intimate tie-up between TI phase and time-reversal symmetry is presently subject of intense experimental and theoretical efforts to understand the response of TSSs to time-symmetry violating magnetic exchange fields $\vec{H}_{\text{Ex}}$ introduced by magnetic guest atoms. For $\vec{H}_{\text{Ex}}$ oriented perpendicular to the surface plane in a stable manner, a gap at the Dirac point (DP) should appear as a signature of losing TSS properties~\cite{Liu2009, Henk2012}.\\
\indent In search for model systems, local spin and orbital moment formation, magneto-crystalline anisotropy of 3$d$ bulk and surface guest atoms, and charge doping were studied particularly for prototypical bismuth chalcogenide TI host materials~\cite{Wray2011, Honolka2012, Scholz2012, Eelbo2014}, where the non-vanished average exchange fields $\left<\vec{H}_{\text{Ex}}\right>$ can be stabilized e.g. by carrier-mediated exchange interactions $J_{\text{3D/2D}}$ between guest atoms~\citep{Biswas2010, Abanin2011, Efimkin2014}.
Dilute surface adatoms are promising since direct experimental access to structural and electronic properties using scanning tunneling microscopy (STM) and angle-resolved photoemission spectroscopy (ARPES) is retained.
Here, ARPES reported intact TSSs~\cite{Scholz2012}, according to a simple picture of 
thermally fluctuating local moments ($J_{\text{2D}}<< k_{\rm B} T$) giving $\left<\vec{H}_{\text{Ex}}\right> = 0$ on the time scale required for a spin-flip process. In contrast, a recent low temperature scanning tunneling spectroscopy (STS) study of local TSS scattering processes claims that paramagnetic but weakly coupled 3$d$ adatom moments are able to locally produce exchange fields on transient time scales sufficient to gap the TI phase~\cite{Sessi2014}. A full understanding is complicated by the presence of the adatoms' non-magnetic resonant scattering potential $U$, which, like in other Dirac materials such as graphene and $d$-wave superconductors, can counteract a moment-induced gap formation at the DP~\cite{Black2015, Sessi2016}.\\
\indent In this paper we report on an efficient quenching mechanism of single Ni adatom moments on Te-terminated Bi$_2$Te$_2$Se and Bi$_2$Te$_3$ surfaces, despite a partially filled Ni 3$d$-shell with occupancy $n_d = 9.2$. Our {\it ab initio} calculations explain this surprising behaviour in the framework of a local Stoner criterion, which predicts Fe adatoms to form large spin moments of $m_s=2.54\,\mu_{\rm B}$ with out-of-plane anisotropy, while Ni remains below an effective Stoner threshold for local moment formation. 
With the DP remaining in the Bi$_2$Te$_2$Se band gap for both elements Ni and Fe, magnetism dependent scattering processes can be ideally studied in this system in the TSS-dominated transport regime, both in macroscopic transport measurements~\cite{Barreto:2014} and on the atomic scale via STS~\cite{Sessi2014}.\\
\indent Our results are based on a thorough determination of structural, electronic and magnetic properties of Ni and Fe adatoms on Bi$_2$Te$_2$Se and Bi$_2$Te$_3$, combining local STM~\cite{Bianchi:2012b,Loeptien2014} with X-ray magnetic circular dichroism (XMCD) spectroscopy (X-Treme beamline, SLS~\cite{Piamonteze2012}) and ARPES (SGM3 beamline at ASTRID2 synchrotron radiation facility~\cite{Hoffmann2004}).
All experiments were carried out with identical Bi$_2$Te$_3$ and Bi$_2$Te$_2$Se single crystals used in Refs.~\cite{Michiardi:2014} and \cite{Mi2013}, respectively. 
For the latter crystals, four-point probe measurements have proven surface-dominated transport with a high bulk resistivity at low temperatures~\cite{Barreto:2014}. 
The crystals were cleaved in ultrahigh vacuum at room temperature using scotch tape followed by cooling down to temperatures below 100~K within $25$~min for ARPES, $20$~min for XMCD, and within $5$~min for STM measurements. Dilute amounts of Ni and Fe in the range of a percentage of a monolayer (ML, number of atoms per surface unit cell) were subsequently deposited from thoroughly outgassed and calibrated e-beam evaporators onto the cold substrate ($T=2$~K for XMCD, $T=7$~K for STM and $T\approx90$~K for ARPES), in order to prevent clustering or subsurface diffusion of Ni and Fe atoms~\cite{Schlenk:2013}.\\
\begin{figure}
\center 
\includegraphics[width=85mm]{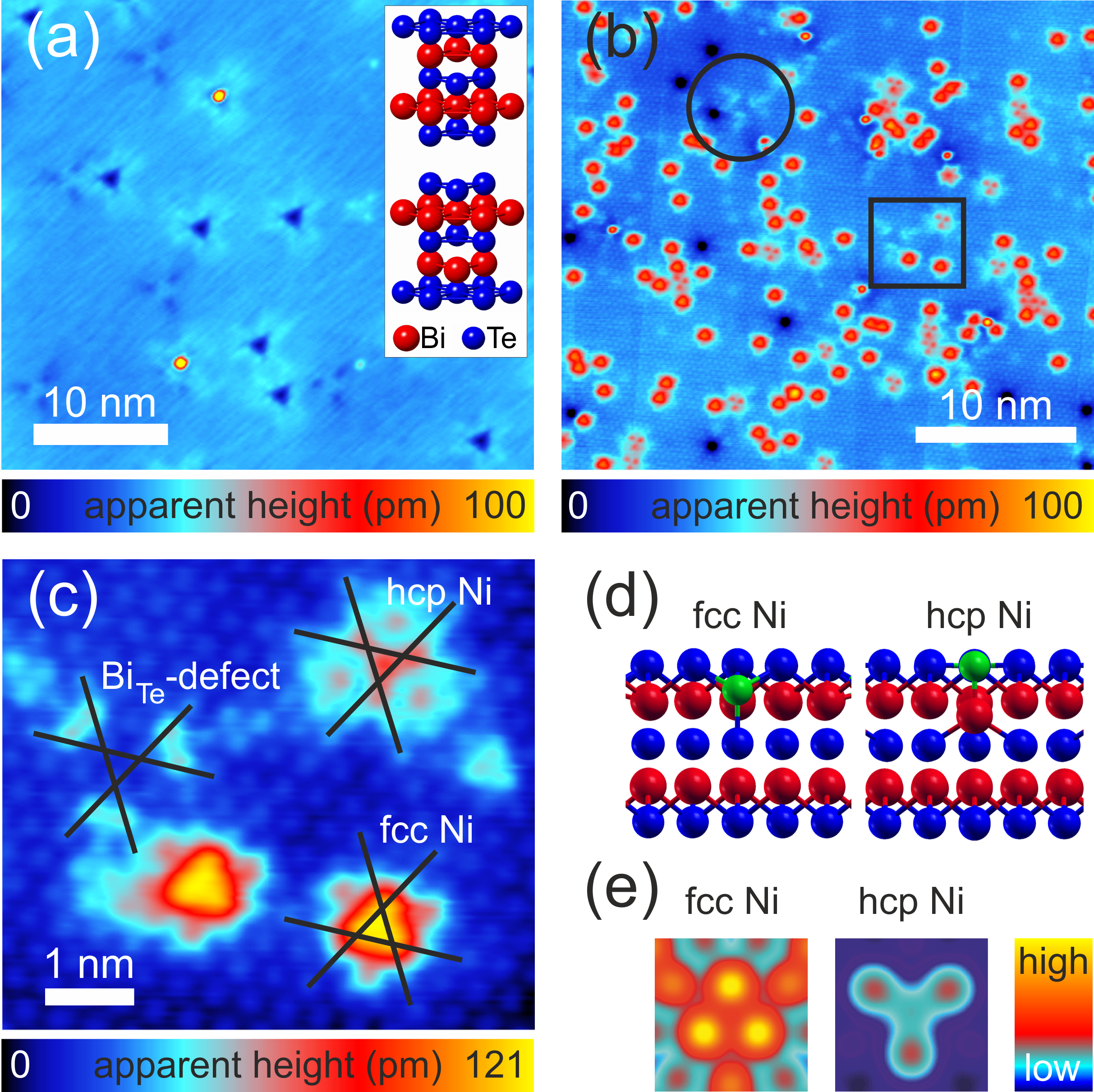}
\caption{
(color online). (a)~STM topograph of the pristine $\mathrm{Bi_2Te_3}$ surface ($U=350$\,mV,  $I=20$\,pA). The inset illustrates two QLs of the crystal structure. (b)~Overview image of $\approx 1\,\%$~ML Ni (red dots and clover shaped defects) on $\mathrm{Bi_2Te_3}$ ($U=-995$\,mV,  $I=400$\,pA). The black circle highlights a typical $\mathrm{Bi_{Te}}$ antisite defect. (c)~Atomically resolved image of the area marked by the black square in (b) with single Ni atoms occupying two different adsorption sites and a $\mathrm{Bi_{Te}}$ antisite defect ($U=-350$\,mV,  $I=400$\,pA, composite of original and Fourier filtered topograph in order to enhance atomic contrast). The lines trace the surface Te atom rows. (d)~Side views of the relaxed positions of $\mathrm{Ni_{fcc}}$ and $\mathrm{Ni_{hcp}}$ from DFT-calculations. (e)~Calculated STM topographs of $\mathrm{Ni_{fcc}}$ and $\mathrm{Ni_{hcp}}$ 
($U=-400$\,mV). For both stackings, the $z$-scale covers the same range. All the STM measurements were performed at $T=4$~K.
}
\label{fign1}
\end{figure}
Figure~\ref{fign1}(a) shows an STM topograph of the pristine $\mathrm{Bi_2Te_3}$ surface after cleavage, which happens at the van der Waals bonds between the quintuple layers (QLs) (see inset) leading to a Te terminated surface. It shows the well known defects in the Te layers of the topmost QL ~\cite{Sessi2013, Bathon2015}. Similar studies on Bi$_2$Te$_2$Se crystals (see Supplemental Material [SM]) show $12\% \pm 3\%$ of $\mathrm{Se_{Te}}$ antisite defects in the surface Te layer. Apart from those infrequent $\mathrm{Se_{Te}}$ defects, the sequence of the topmost two layers of $\mathrm{Bi_2Te_3}$ and Bi$_2$Te$_2$Se is thus identical and we expect similar adsorption geometries and magnetic properties for the $3d$ atoms deposited on both Te-terminated substrates.
Figure~\ref{fign1}(b) shows the STM topograph after deposition of 1$\%$~ML Ni. In addition to the features from the intrinsic defects in the topmost Te layers (black circle), there is a statistical distribution of defects stemming from Ni adatoms which have an apparent height of $25$\,pm - $110$\,pm depending on the bias voltage. Closer inspection reveals that the Ni atoms have two distinct appearances due to adsorption on either of the two hollow sites of the surface Te lattice (Fig.~\ref{fign1}(c), see lines). Considering that the clover-shaped subsurface defects are attributed to $\mathrm{Bi_{Te}}$ antisites residing in the fifth layer (Te) of the first QL ~\cite{Wang:2011, Bathon2015} and their position relative to the surface Te atoms (see lines), we can assign the shallower Ni atom to $\mathrm{Ni_{hcp}}$ and the higher one to $\mathrm{Ni_{fcc}}$. They have a strongly different relative abundance $\mathrm{Ni_{hcp}}$:$\mathrm{Ni_{fcc}}$ of 0.2. For Fe on $\mathrm{Bi_2Te_3}$, our STM experiments confirm the previously revealed hollow site adsorption, appearance in STM and relative abundance $\mathrm{Fe_{hcp}}$:$\mathrm{Fe_{fcc}}$ of 0.7~\cite{Eelbo2014}. Very similar results are found for the adsorption of the two elements on Bi$_2$Te$_2$Se (not shown).\\
In order to determine the vertical position of Ni for the electronic structure calculations shown below, we performed DFT calculations equivalent to those in Ref.~\cite{Honolka2012}. We started from the two lateral positions of the adatoms revealed by STM (Fig.~\ref{fign1}(c)) and then relaxed them (Fig.~\ref{fign1}(d)). While $\mathrm{Ni_{fcc}}$ strongly relaxes into the first Bi layer, $\mathrm{Ni_{hcp}}$ sits in the surface Te layer and pushes the underlying Bi atom downwards. Using these two adsorption geometries we calculate STM images  (Fig.~\ref{fign1}(e)) with the Tersoff-Hamann approach~\cite{Tersoff1984} by evaluating the partial charge density in a range of 0.4~eV below the Fermi energy ($E_{\rm F}$) at $\mathrm{\sim2.1~\mathring{A}}$ above the surface. The images confirm the experimentally observed difference in height and appearance of $\mathrm{Ni_{fcc}}$ and $\mathrm{Ni_{hcp}}$ as well as the assignment to the adsorption sites. Our calculations for the case of Fe confirm previous results \cite{Eelbo2014}.\\
\begin{figure}
\center 
\includegraphics[width=85mm]{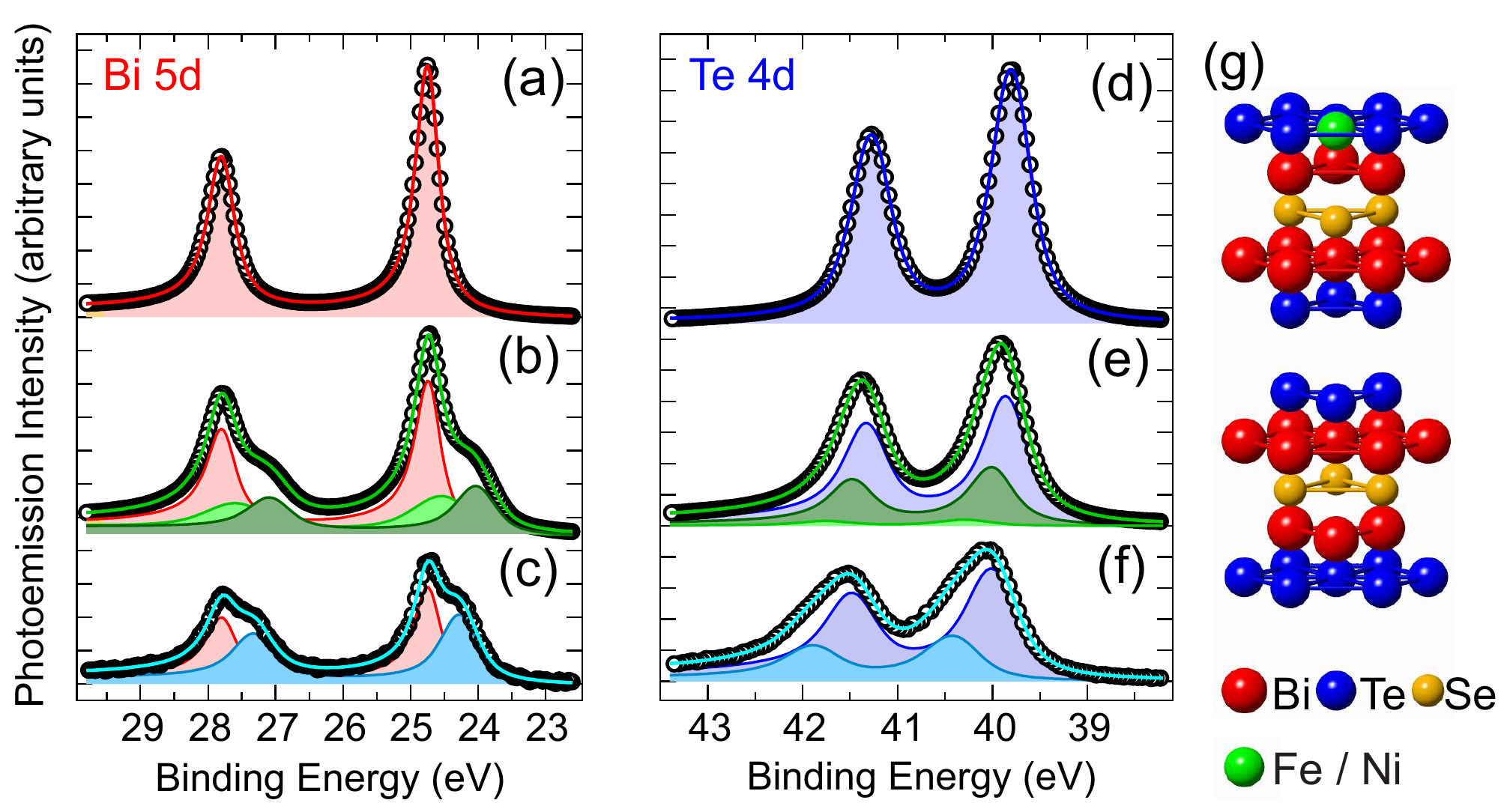}
\caption{
Bi~5$d$ (left panel) and Te~4$d$ (right panel) XPS core level spectra at photon energy $h \nu = 100\,$eV and $T\approx90$~K for pristine Bi$_2$Te$_2$Se [(a), (d)], as well as after deposition of 0.4~ML~Fe [(b), (e)] and 0.5~ML~Ni [(c), (f)], respectively. Fits represent decompositions into Bi and Te doublets. (g) Illustration of the top two QLs of Bi$_2$Te$_2$Se with an adsorbed adatom.
}
\label{fign2}
\end{figure}
\begin{figure}
\center 
\includegraphics[width=85mm]{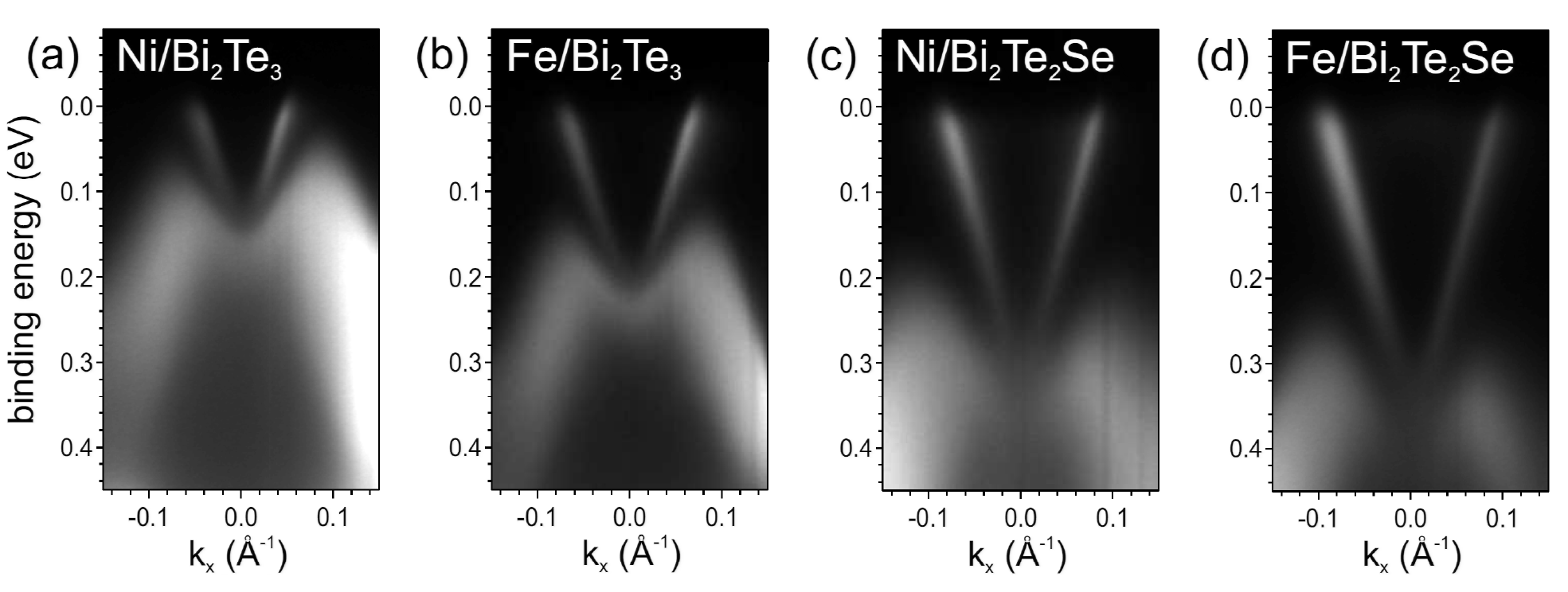}   
\caption{
ARPES data for $\approx0.3\%\ $ML\ Ni on $\mathrm{Bi_2Te_3}$ (a) and Bi$_2$Te$_2$Se (b), respectively, and $\approx0.2\%\ $ML\ Fe on $\mathrm{Bi_2Te_3}$ (c) and Bi$_2$Te$_2$Se (d), respectively. Spectra were taken at photon energy $h \nu = 21\,$eV and $T\approx90$~K.
}
\label{fign3}
\end{figure}
\begin{figure*}
\center 
\includegraphics[width=16.5cm]{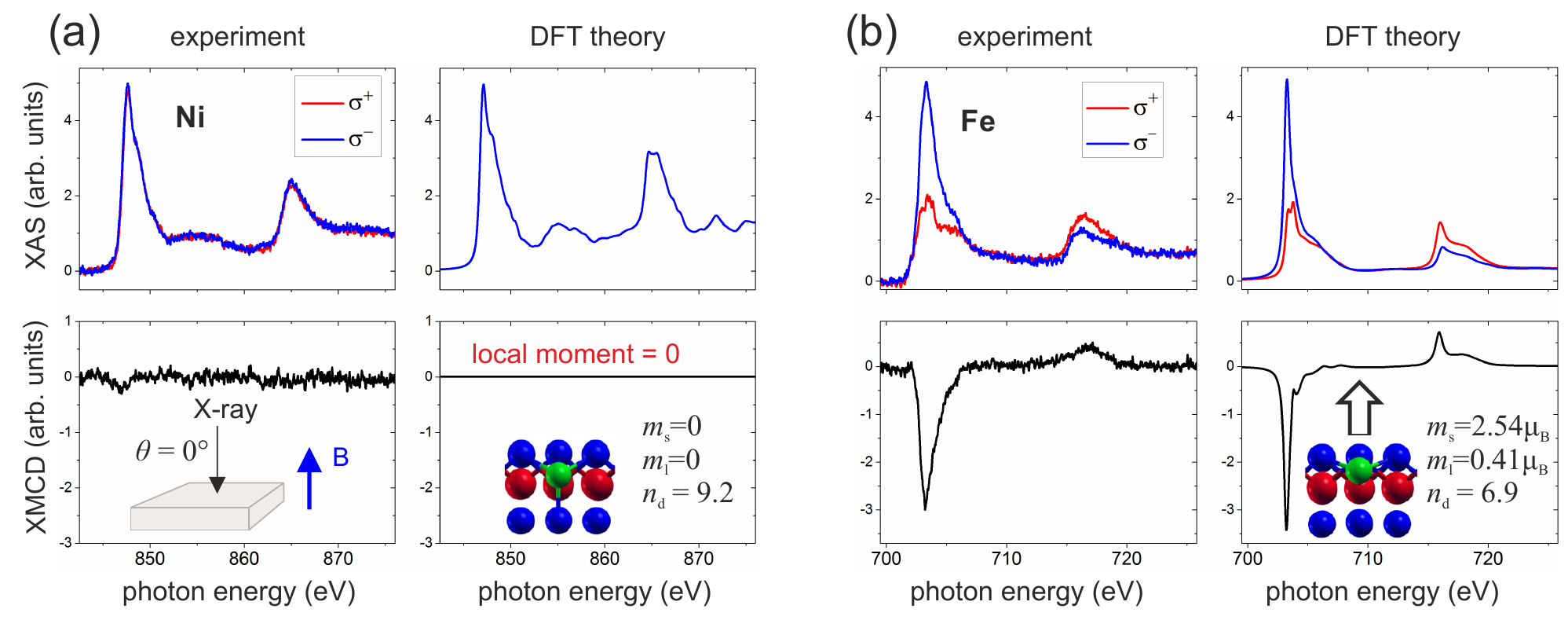}
\caption{
(a) and (b): Ni and Fe L$_{2,3}$ XAS data on Bi$_2$Te$_2$Se for positive ($\sigma^+$) and negative ($\sigma^-$) circular polarization in polar geometry ($\theta = 0^{\circ}$) and resulting XMCD signals below (left panels). Measurements were done in the impurity regime ($\approx 1\,\%$ coverage) at low temperatures of $T=2$\,K and polar fields of $B=6.8\,$T. Right panels show respective spectra calculated for $\mathrm{Bi_2Te_3}$ on the basis of DFT theory using the relaxed fcc geometry shown as insets. Ni and Fe both show partially filled $d$-shells but only Fe develops finite local orbital and spin moments.
}
\label{fign4}
\end{figure*}
The top Te-Bi layer is identical for $\mathrm{Bi_2Te_3}$ and Bi$_2$Te$_2$Se (see structural models in Figs.~\ref{fign1}(a) and ~\ref{fign2}(g) ~\cite{Jia:2012}), suggesting similar relaxed positions of Ni or Fe. As a consequence we expect the same chemical bonding and magnetic properties for those substrates, as confirmed by X-ray photoemission spectroscopy (XPS) and XMCD results (see below).\\
Surface sensitive XPS of the Bi~$5d$ and Te~$4d$ core levels from pristine Bi$_2$Te$_2$Se in Figs.~\ref{fign2}(a,d) develop strong shoulders upon deposition of 0.4~ML Fe (b,e) and 0.5~ML Ni (c,f), while the Se~$3d$ core level spectrum does not show significant changes in the line shape. This is consistent with metal adatoms residing between the two topmost layers, where they cause strong chemical shifts in neighboring Te and Bi atoms, but not in Se (see Fig.~\ref{fign2}(g)), because adatoms have little influence on the third subsurface layer. Ni and Fe similarly shift Bi~$5d$ (Te~$4d$) lines to lower (higher) binding energies, in accordance to a weakening of the polar Bi-Te bond towards a more metallic bonding, as expected. The peak decomposition (SM) shown in Fig.~\ref{fign2} reveals, that, while Ni deposition generates a single component shifted by 0.46~eV, Fe undoubtedly forms two distinct components shifted by 0.23~eV and 0.72~eV with respect to the main pristine Bi doublet. Although coverages here are beyond the impurity regime, these two components are consistent with the STM results which showed that Fe atoms sit directly above Bi atoms (hcp site, stronger chemical shift) and above empty spaces (fcc site, weaker chemical shift) with nearly equal probability. Ni atoms in contrast strongly prefer fcc sites leading to only one shifted component. Te spectra in Fig.~\ref{fign2}(b) show only one additional component in both cases Ni and Fe, very likely because both fcc and hcp adatoms are embedded in a Te triangular geometry.\\
\indent In order to study the doping effect of adatoms on the TSS, we performed ARPES for different coverages. Generally, the surface impurities lead to a downwards energy shift of the TSS and bulk bands~\cite{Jia:2012}, which saturates at coverages in the percentage range of a ML. This indicates electron transfer from the adatom to the substrate. As shown in Fig.~\ref{fign3}, at $\leq 1\,\%$ coverages, only the linearly dispersing TSS crosses $E_{\text{F}}$ located in the bulk band gap of Bi$_2$Te$_3$ and Bi$_2$Te$_2$Se. Thus, these systems represent the very particular case where transport will be determined exclusively by the interaction of the TSS with the localized 3$d$ guest atoms charge and magnetic moment, which will be investigated in the following.\\
\indent The late 3$d$ transition element Ni is known to be a borderline element for transitions between high-spin $3d^{8}4s^{2}$ and low-spin $3d^{9}4s^{1}$ configurations already in the free atom case or on weakly interacting graphitic substrates~\cite{Godehusen2002, SessiV2014}, and thus adatom electronic and magnetic ground state configurations are highly sensitive to interactions with the host surface. For an analysis of the magnetic ground state properties on Bi$_2$Te$_3$ and Bi$_2$Te$_2$Se we performed X-ray absorption spectroscopy (XAS) at the L$_{3,2}$ edges (Fig.~\ref{fign4}), which provide information on the local electronic state, since spectral shapes and resonant absorption intensities depend on the distribution of holes in the 3$d$ shell. For magnetic contrast we applied large magnetic fields of $B=6.8$~T collinear to a positive (+) and negative (-) circularly polarized X-ray beam, as depicted in Fig.~\ref{fign4}(a). Differences in the respective absorption spectra $\sigma^+$ and $\sigma^-$ reflect the magnetic ground state of single 3$d$ atoms as described in an earlier work~\cite{Honolka2012}. Fig.~\ref{fign4}(a) and (b) summarizes Ni and Fe spectroscopy in the single impurity limit and corresponding DFT results based on the spin-polarized relativistic multiple-scattering or Korringa-Kohn-Rostoker formalism using coherent potential approximation~\cite{EKM11} together with Fermi's golden rule~\cite{ME04}. All XAS data in Fig.~\ref{fign4} were measured at normal incidence $\theta = 0^{\circ}$ in the total electron yield mode, and spectra are shown after Bi$_2$Te$_2$Se background subtraction. Before and after XAS characterization, the samples were carefully checked for oxygen contaminations using the XAS at the O K-edge.\\ 
\indent Prominent resonant multiplet structures visible especially at the leading L$_{3}$ edge of Ni and Fe spectra are typical for partly filled localized $d$-shells weakly hybridized with their environments. The well-resolved multiplet structures in XAS serve as a fingerprint of the local $d$ shell configurations and show an exceptional agreement with the ones derived from DFT (right panels of (a) and (b)).
They correspond to unfilled Ni and Fe $d$-shell occupancies $n_d=9.2$ and $n_d=6.9$, respectively.  Strikingly, while Fe develops a local moment as reflected by the Fe XMCD signal ($\sigma^+ - \sigma^-$), Ni takes a non-magnetic ground state (see lower panels in Fig.~\ref{fign4}(a) and (b)). Corresponding spectroscopy data at grazing incidence $\theta = 60^{\circ}$ gave similar results for Ni, while Fe spectra showed a pronounced out-of-plane magnetic easy axis with a magnetic anisotropy of $7 \pm 3\,{\rm meV/atom}$ (see SM) in agreement with Eelbo et al.~\cite{Eelbo2014}. Consistently, the DFT calculations result in local moments of $m_s=0\,\mu_B$ and $m_l=0\,\mu_B$ for Ni and $m_s=2.54\,\mu_B$ and $m_l=0.41\,\mu_B$ for Fe. This quenching of the Ni atomic moment is a consequence of the interaction of the 3$d$ element with the top Te-Bi layer as the experimental finding is the same for binary Bi$_2$Te$_3$ (see SM).\\
\indent To gain an understanding of the mechanism responsible for the local Ni moment quenching in a partially filled 3$d$ shell, we compare the stability of the magnetic states of relaxed Ni and Fe atoms with our DFT calculations. The local densities of states (DOS) in Fig.~\ref{DOS} show that despite of the localized character of Ni and Fe $d$-electrons, the width of the partially occupied energy bands is rather large due to hybridization with the electronic states of surrounding host atoms and comparable to the width of energy bands in 3$d$ itinerant ferromagnets. This suggests to use a Stoner criterion $(I\chi_{0})^{k}\geq 1$ for the spontaneous formation of a spin moment on atom $k$ (Ni or Fe in our case), where $(I\chi_{0})^{k}$ is the so-called Stoner product to analyze the atom of type $k$. For the most simple situations  $(I\chi_{0})^{k}$ can be split into the Stoner integral $I^{k}$ and the enhanced spin susceptibility $\chi_{0}^{k}=2\mu_{\rm B}n^{k}(E_{\rm F})$, where $n^{k}(E_{\rm F})$ is the DOS at $E_{\rm F}$. For the more general case of an alloy and adatoms considered here, we express $(I\chi_{0})^{k}$ using linear response theory as suggested by Deng et al.~\cite{DFVE01} (see SM). 
For Fe atoms, our self-consistent calculations give a Stoner product $(I\chi_0)^{\text{Fe}}=2.68$ well above 1, while for Ni we find only $(I\chi_0)^{\text{Ni}}=0.78$. We thus can understand the quenching mechanism on the atomic level surprisingly well within the Stoner model, triggered by a critically low DOS at $E_{\rm F}$ on the Ni site close to the edge of Ni $d$-band (Fig.~\ref{DOS}).\\
In summary, we report on the complete quenching of Ni adatom moments sitted on Te-terminated bismuth chalcogenide topological insulator surfaces. Comparing adatom site and electronic properties of Ni with magnetic Fe, our study allows understanding the quenching in the framework of a local Stoner theory. We find Fermi levels in the gap of Bi$_2$Te$_2$Se in the presence of adatoms, perfect for investigating topological insulators under nonmagnetic Ni and out-of-plane Fe moments perturbation.

\begin{acknowledgments}

The authors acknowledge financial support from the Czech Science Foundation (grants No.\ GA13-30397S and GA14-30062S), 
from Ministry of Education, Youth and Sports of the Czech Republic (grant LM2011029 and LO1409), 
from the Deutsche Forschungsgemeinschaft (DFG) through SPP 1666 (under project Eb-154/26 and WI 3097/2-2), 
from the Aarhus University Research Foundation and the VILLUM foundation,
from COST Action MP1306 (EuSpec) and CENTEM PLUS (LO1402).
In addition JWa and AAK acknowledge support from the DFG via the Emmy Noether Program (KH324/1-1) 
and JH acknowledges the Purkyn\v{e} Fellowship program of the Czech Academy of Sciences.
The X-ray absorption measurements were performed on the EPFL/PSI X-Treme beamline at the Swiss Light Source, Paul Scherrer Institut, Villigen, Switzerland. The research leading to these results has received funding from the European Community's Seventh Framework Programme (FP7/2007-2013) 
under grant agreement n.°312284.

\end{acknowledgments}

\bibliography{Bi2Se3library}


\end{document}